\documentclass[runningheads]{llncs}

\usepackage[T1]{fontenc}
\usepackage{graphicx}
\usepackage{graphicx}
\usepackage{booktabs} 

\usepackage{graphicx}
\usepackage{booktabs} 
\usepackage{enumitem}
\usepackage[ruled]{algorithm2e} 

\SetAlFnt{\small}
\SetAlCapFnt{\small}
\SetAlCapNameFnt{\small}
\SetAlCapHSkip{0pt}
\IncMargin{-\parindent}
\usepackage{amsmath}
\usepackage[utf8]{inputenc} 
\usepackage[T1]{fontenc}    
\usepackage{hyperref}       
\usepackage{url}            
\usepackage{booktabs}       
\usepackage{amsfonts}       
\usepackage{nicefrac}       
\usepackage{microtype}      
\usepackage[capitalise,nameinlink]{cleveref}
\usepackage{xspace}
\usepackage{xcolor}

\newcommand{\eps}{\ensuremath{\epsilon}\xspace}

\newcount\Comments  
\definecolor{darkgreen}{rgb}{0,0.6,0}
\newcommand{\kibitz}[2]{\ifnum\Comments=1{\color{#1}{#2}}\fi}

\newtheorem{assumption}{Assumption}

\crefname{assumption}{Assumption}{Assumptions}

\begin{document}
\title{Learning Approximately Optimal Contracts}
\author{Alon Cohen\inst{1} \and
Argyrios Deligkas \inst{2} \and
Moran Koren\inst{3}}
\authorrunning{A.Cohen et al.}
%
\institute{Tel-Aviv University and Google Research \email{alonco@tauex.tau.ac.il}\and
Royal Holloway, University of London 
\email{Argyrios.Deligkas@rhul.ac.uk}\and
Tel Aviv University
\email{me@mkoren.org}
}

\maketitle

\begin{abstract}
In  principal-agent models, a principal offers a contract to an agent to preform a certain task. The agent exerts a level of effort that maximizes her utility. The principal is oblivious to the agent's chosen level of effort, and conditions her wage only on possible~outcomes. 
In this work, we consider a model in which the principal is unaware of the agent's utility and action space: she sequentially offers contracts to identical  agents, and observes the resulting outcomes. 
We present an algorithm for learning the optimal contract under mild assumptions. We bound the number of samples needed for the principal obtain a contract that  is within \eps of her optimal net profit for every $\eps>0$. Our results are  robust even when considering risk averse agents.  
Furthermore, we show that when there only two possible 
outcomes, or the agent is risk neutral,  the algorithm's outcome approximates the optimal contract described in the classical theory.
\end{abstract}



\section{Introduction}

Recent technological advances have changed the relationship between firms and employees dramatically. The rapid change in employees' required skill set combined with the availability of off-shore, qualified, cheap hiring alternatives drove employers to adopt employees on a short-term, task specific bases. This hiring model, dubbed \textit{dynamic workforce} or \textit{gig-economy}, has been growing exponentially over the decade. In \cite{Bughin2016} it was found that 20-30\% of US workers are employed independently, at least partly, and predicts this trend will continue with popularity of the ``Lean Startup'' business model, and the appearance of platforms such as ``Amazon Mechanical Turk''.

The aforementioned  labor market changes have a profound effect on the information  available to firms when it comes to making hiring decisions.
In the traditional workspace, hiring is often a long-term, expensive procedure. Nowadays hiring is cheap and short-termed. As a result, an employer has less information on the character and quality of her employees, thus lacking the knowledge on how to best align their incentives with those of the firm. In this work we study a model more suitable to the new ``dynamic workforce'', and, using this model, we try to learn the best wages an employer should offer to her employees.

Economists use the term {\em agency problems} to describe models like the above. In such models one party, the {\em principal}, offers a {\em contract} to another party, the {\em agent}, to perform a task~\cite{laffont2009}.  
In his seminal paper, Ross~\cite{Ross1973} introduced the term {\em principal--agent} 
to  capture these models. In the basic model, the outcome of the task is chosen randomly from a distribution determined by the {\em level of effort} invested by the agent. A higher effort level induces a distribution in which the probability of a better outcome is higher than in lower effort levels. On the other hand, a higher level causes the agent greater disutility.    
Ross~\cite{Ross1973} assumes that the principal is {\em risk-neutral} while the agent is {\em risk-averse}, therefore the incentives of both parties are {\em not} fully aligned.\footnote{When given the choice between participating in some lottery $X$ or receiving $E(X)$ with probability one, a risk-averse agent will strictly prefer the latter while a risk-neutral agent is indifferent. In the Von Neumann Morgenstern utility theory, if an agent is risk-averse, then she has a concave utility function; if she is risk-neutral, then her utility is linear.} Hence, the effort level that is optimal from the perspective of the principal, is not necessarily the optimal one for the agent. To bridge this gap, the principal offers the agent a contract in which the agent is rewarded for any additional effort. 
The lion's share of previous work assume that the principal, while oblivious to the agent's choice, has full information about the agent's utility structure {\em and} the set of levels of effort she can choose from {\em and} the probability distribution associated with every level of effort \cite{Ross1973,Grossman1983,Holmstrom1979,Gershkov2012,Holmstrom2017}. These assumptions seem confining, especially when considering the motivating scenario of a dynamic workplace.

Holmstrom, in his Nobel Prize lecture~\cite{Holmstrom2017}, identified this gap in theory as one of the main challenges in current research.
A first attempt to bridge this gap was \cite{Carroll2015}. 
Carroll \cite{Carroll2015} presented a model where a principal has only partial knowledge on the agent's action space. He assumed that the agent is risk neutral and thus, if the principal have known the agent's complete action space, the optimal contract would have been linear. 
In this setting, the principal can guarantee the net profit from best linear contract pertaining the actions she knows. That is her actual net profit will never be lower.  
An early attempt to approach the agency problem from the lens of theoretical computer science was made in \cite{Ho2016}.
In the model of~\cite{Ho2016}, there are several types of agents, each one with her own utility function, set of effort levels and effort costs, and probability distributions over the outcomes; all of which unknown to the principal. They considered a repeated setting with $T$ rounds, where in every round the type of the agent is chosen i.i.d. from an unknown probability distribution. 
Their goal was to find, before round $T$, the optimal contract from a predefined finite set $S$. 
They followed a multi-armed bandit approach~ \cite{robbins1985some} and derived an algorithm that finds an approximately-optimal contract. However, the contract they find is approximately-optimal {\em only} with respect to the best contract in $S$. They provide no theoretical guarantees on approximating the optimal contract overall, but only for the case of one effort level for the agent, i.e. reject the contract or accept it, known as {\em posted-price auctions}.\footnote{ As they highlight, it is not generally clear whether the best contract from $S$ can provide a good theoretical guarantees for the general problem of dynamic contract design.} 
In  \cite{DuttingRT19-simple}, the authors study a similar scenario in which the principal knows only the expected output generated by each contract, that is, the agent's action set is unknown to the principal. As in \cite{Carroll2015}, they assume that the agent's utility is linear and that limited liability holds (i.e., no negative payments) and show that  the worst-case expected profit for the principal is guaranteed by a linear contract. In addition, they provide tight approximation guarantees for linear contracts.   
In this paper we follow a similar route to~\cite{Ho2016} and \cite{DuttingRT19-simple}. We assume the principal has {\em zero} information about the agent; we assume unknown utility for the agent, unknown set of effort levels and their associated costs, and unknown probability distribution for any effort level.  The only knowledge the principal has is the set of outcomes and their corresponding profits. Our model extends the setting of \cite{Ho2016} as we assume that the contract set as continuous. We extend the setting of \cite{Carroll2015} and \cite{DuttingRT19-simple} in two folds: (1) Our set of {\em monotone-smooth} contracts is richer than the set of linear contracts. (2) We do not assume any functional form of agent utilities, we only assume that agents are risk averse (concave utility). In the next section we illustrate the importance of this generalization in this context.

 \subsection{The challenge of learning contracts with risk averse agents.}\label{sec:ra}

In contract theory, we study the tension between a principal and an agent. This tension is the result of misalignment between the incentives of both players. On the one hand, the principal wishes to maximize her profit, and is indifferent towards the cost the agent endures during effort; on the other hand, the agent is assumed to be risk averse, and wishes to minimize the volatility of her expected wage (ceteris paribus). In the classical PA model, the principal knows the full structure of the agent's decision problem, and can offer a contract that serves her best yet minimizing the aforementioned misalignment with the agent.  To illustrate this, we show an example where the agent may choose between three levels of effort and there are two possible outcomes. As we wish to illustrate a general point, we will keep the exact utility function and cost structure implicit and just mention that the utility is concave and higher costs induce ``better" output distributions. The  following figure depicts the space of contracts from which the principal may offer, and the resulting effort level the agent chooses, given each possible contract.  Note that unlike \cite{Ho2016}, the space of contracts is continuous.
\begin{figure}[!h]
\centering\includegraphics[width=0.7\textwidth]{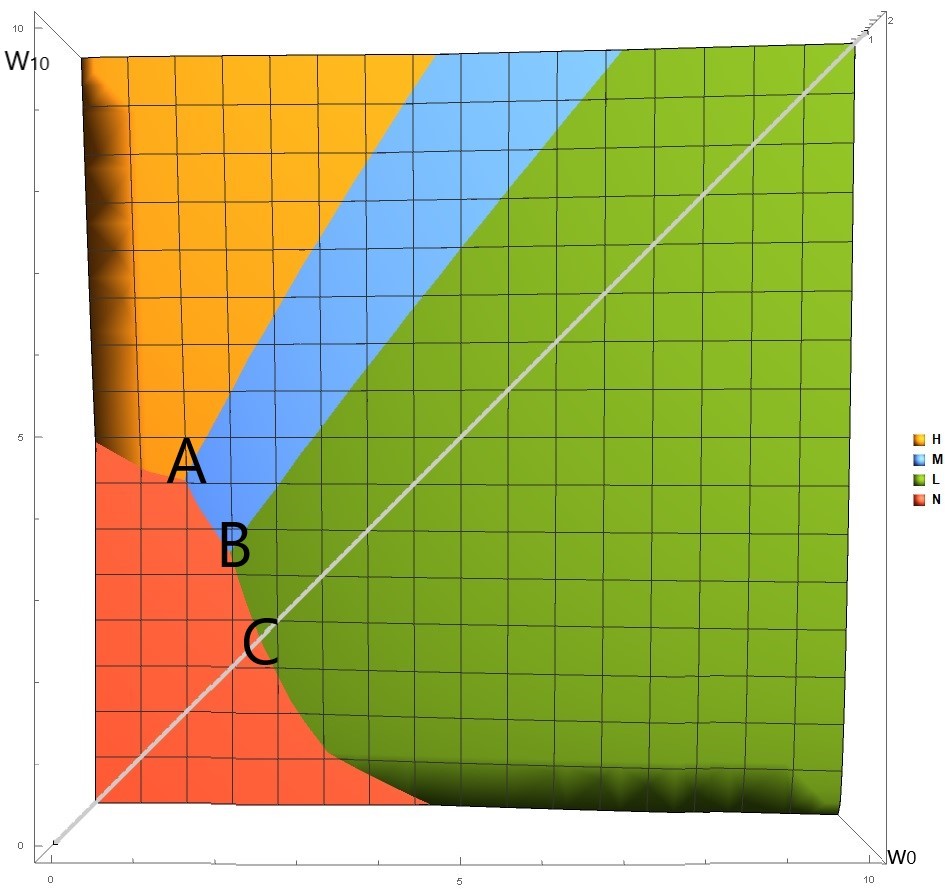}\caption{The agent's optimal choice in a PA model with two possible outputs, three levels of effort:$e_H,e_M,e_L,$ $N$ is the rejection zone, and a concave utility. }
\end{figure}
By \cite{Mas-Colell1995}, the optimal contract can be found in any of the interaction points $\{A,B,C\}$. To see why this is the case, note, for example, that when offered the contract in point $A,$ the agent is indifferent between accepting and rejecting the contract, and between investing medium effort or high effort.  Thus, based on the model assumptions, she will choose to invest the high level of effort. Not only that, but this is the contract where the highest level of effort is achieved for the lowest possible expected pay. 
Similarly, the contract $B$, is the ``cheapest'' contract under which the agent chooses the middle effort level.  Note that contract $B$ is both cheaper, and yields lower expected output than contract $A.$ When examining the principal-optimal contract, it is unclear which of the effects dominates. Therefore, in the general case, the optimal contract can be in any of $A,B,C.$ When the principal is fully aware of the agent's incentive structure (i.e., her utility and effort-cost pairs), finding the optimal contract is straightforward.

If we consider  an uninformed the principal, she can attempt to learn an optimal contract using the following pesudo-algorithm: (1)  create a discretized version of the contract space (2) Find the contract on the grid which yields the optimal profit.  Naively, using this approach, we will find a contract that is at most $\varepsilon$ away either $A, B$ or $C$.  Note that this does not preclude the possibility that the contract that we find is $\varepsilon$ away from point $A$, but the absolute optimal contract is actually contract $B.$ In extreme cases, the difference in the principal's profit between the algorithms' result and the theoretically optimal one can be arbitrarily high (see for example $u(w)=1-\frac{1}{w}).$ Furthermore, using a finer grid may lead to different contract which is arbitrarily away, thus convergence of the process in not guaranteed.  In this paper, we provide a set of contracts, and algorithm, and an appropriate discretization for which the process converges to a single maximal contract. In addition, we detail two cases, for which the resulting contract is the absolute possible best.

\paragraph{\bf Further related work.} 
As we have already explained, the majority of works so far has focused on the full information setting and its variants:~\cite{babaioff2012combinatorial-agency} studies a model with many agents;~\cite{babaioff2010mixed} studies a setting with externalities for the agent;~\cite{DuttingEFK21-combinatorial,DuttingRT21-complexity} study  principal-agent models with combinatorial structures over the actions of the agent, or the possible outcomes.  Additionally, in \cite{DuttingRT21-complexity}, the authors also discuss a black-box model and present sample-complexity results. 
More recently, a new line of work on Bayesian settings for the principal-agent model has emerged~\cite{alon2021contracts,castiglioni2022bayesian,castiglioni2022designing,guruganesh2021contracts} where the agent type comes from a known probability distribution.

Our work lies in the intersection of principal-agent models %
and multi-armed bandit theory; 
\cite{Ho2016} provide an excellent overview of literature in the field.
\cite{Sannikov08,Sannikov11,williams2004} study a repeated setting where the principal interacts with the agent for multiple periods.
\cite{Conitzer2006} empirically compare several learning algorithms in a setting similar to~\cite{Ho2016}. 

The Lipschitz Bandit problem \cite {agrawal1995continuum} is a generalization of the multi-armed bandit problem in which the set of arms comes from some compact space, and the expected reward of each arm is a Lipschitz function of the arm. It has received much attention from the bandit theory community over the years \cite{kleinberg2005nearly,auer2007improved,bubeck2011lipschitz,kleinberg2013bandits,magureanu2014lipschitz}. 



\paragraph{\bf Our contributions.}

In this work we make several contributions to both economic and computer science theory. 
\begin{itemize}[nosep,leftmargin=*]
	\item We contribute to the study of agency problems by extending the literature to a setting where the principal is oblivious to the agent decision problem structure.
	\item We study this setting in a the context of risk averse agents. An extension which is the driving force of the canonical theoretical principal-agent model.  To the best of our knowledge, our paper is  first in the  introduction of agent risk aversion into the theoretical study of data driven decision problems. 
	\item We 
	introduce a novel set of contracts we call {\em monotone-smooth} contracts; a large subset of monotone contracts set studied in \cite{Ho2016}. 
	\item We complement this definition with a suitable 
	discretization of the contract space.
	Unlike \cite{Ho2016}, we show that for any monotone-smooth contract 
	there exists a contract in the discretized space for which the principal's expected net profit is 
	$\eps$-approximated. As far as we are aware, this is the first work to do so.
	\item Moreover, our result does not assume any specific agent utility function, but rather only mild assumptions. This allows to apply machineries from multi-armed bandit theory to find a contract $\eps$-optimal against any monotone-smooth contract. 
	\item Finally, we present two fundamental cases in which economic theory suggests the  learned contract of our algorithm is $\eps$-optimal against \emph{any} contract the principal may offer. The first case in when there are only two possible outcomes. The second case
	is when there are many outcomes and the agent is risk neutral.
\end{itemize}


\section{Preliminaries}
In what follows, $[m^\star] := \{0,1,\ldots,m\}$ and $[m] := \{1,\ldots,m\}$ for every natural $m$.

We study principal-agent problems with $k$ outcomes and $n$ effort levels. 
Let $0 < \pi(1) < \pi(2) < \cdots < \pi(k)$ denote the value the principal gets 
under outcome $i \in [k]$. We assume that $\pi(k) = H$, hence the values are bounded.

A {\em contract} $w=(w(1), \ldots, w(k))$ specifies a positive payment to the agent 
for every outcome; namely, $w(i)$ is the wage the principal pays the agent for outcome~$i$.

Upon receiving a contract, the agent chooses an effort level $e \in [n^\star]$. 
Every effort level is associated with a probability distribution $f_e$ over the set of 
outcomes and a cost $c(e)$; $f_e(j)$ is the probability of realizing outcome $j$  when 
the agent chooses the effort level $e$. 
In effort level 0, the agent rejects the contract, her utility is zero under any 
contract, and by convention the value for the principal is zero. 
We assume that the effort levels are ordered, i.e., $1 \prec \ldots \prec n$ (the order will be formally defined in Assumption \ref{assumption:fosd}). 
We follow the literature and assume that the agent has a von Neumann-Morgenstern 
utility. For a contract $w = (w(1), \ldots, w(k))$ the agent chooses effort level of 
$\hat e(w)$ as to maximize her utility, defined as 
\begin{align*}
U(w,e) = \sum_{j=1}^k f_e(j) \cdot u(w(j)) - c(e),
\end{align*}
where $u$ is a monotonically-increasing concave function.
Hence, $\hat e(w) = \arg\max_{e \in [n^*]} U(w,e)$.

The principal is risk-neutral; when she offers contract $w$ to the 
agent, her expected net profit from the contract is,
\begin{align*}
V(w)=\sum_{j=1}^k f_{\hat e(w)}(j) \cdot \big(\pi(j) - w(j) \big)~.
\end{align*}

To ensure that higher effort levels yield higher expected profit for the principal, the literature commonly lays down some assumptions about the outcome distributions. 
\begin{assumption}[First-order Stochastic Dominance (FOSD)]
\label{assumption:fosd}
A probability distribution associated with higher effort {\em first 
order stochastically dominates} a probability distribution associated with lower effort. Formally, if $e \succ e'$ , then for every $j \in [k]$ it holds that $\sum_{i=j}^k f_e(i) \geq \sum_{i=j}^k f_{e'}(i)$.
\end{assumption}
Note that the assumption is equivalent to the following. For every pair of effort levels
$e \succ e'$ and for every sequence of real numbers $a(1) \le \cdots \le a(k)$, 
\begin{equation}
\label{eq:assump}
\sum_{i=1}^k f_e(i) \cdot a(i) \ge \sum_{i=1}^k f_{e'}(i) \cdot a(i)~.
\end{equation}

Additionally, to break ties between effort levels we assume the following.
\begin{assumption}
\label{assumption:breakties}
The agent will choose the higher effort when indifferent between two or more levels of effort.
\end{assumption}

In this work, the principal is faced with a stream of agents. 
The agents are all different but identical---they share a common utility function, effort levels, costs from effort, and outcome distributions associated with each effort level. 
The principal proceeds in rounds $t=1,2,\ldots$. On round $t$, the principal offers a 
contract $w_t$ to the agent associated with this round. The agent privately chooses 
effort level $\hat e(w_t)$ unknown to the principal. The principal observes only the 
outcome $i_t$ independently drawn from $f_{\hat e(w_t)}$, and consequently gets a net 
profit of $\pi(i_t) - w_t(i_t)$.  

In what follows, for $\eps > 0$, the goal of the principal is to find an 
\eps-optimal contract in the minimum number of rounds. A contract $w$ is 
{\em \eps-optimal} if $V(w) \geq V(w') - \eps$ for every $w' \in W$, for 
a set of contracts $W$ to be defined in the sequel.

\subsection{Multi-armed Bandit}

In the multi-armed bandit problem \cite{robbins1985some}, a decision maker sequentially collects rewards from a given set of arms. In each round, the decision maker chooses a single arm, and observes an independent sample from a reward distribution associated with that arm. In our case, the goal of the decision maker is, after a predetermined number of rounds, to select an \eps-optimal arm; that is, an arm whose expected reward is at most \eps less than the expected reward of any arm.

When the set of arms is finite, of size $N$, and the rewards are bounded in $[0,B]$, the seminal work of \cite{even2006action} presents an algorithm called \textsc{MedianElimination} with the following guarantee.
\begin{theorem}[\cite{even2006action}]
\label{thm:me}
The \textsc{MedianElimination}$(\epsilon, \delta)$ returns an \eps-optimal arm with probability at least $1-\delta$ after $O((N B^2/\epsilon^2) \cdot \log(1/\delta))$ rounds. 
\end{theorem}

In our problem, each contract can be seen as an arm. The expected reward of each arm is exactly the principal's utility associated with this contract. It is then expected that the principal would simply execute \textsc{MedianElimination} on the space of contracts to obtain an \eps-optimal one. However, the space of contracts is not finite which is crucial for \textsc{MedianElimination} to run. In the sequel we show how to overcome this difficulty by discretizing the space of contracts, and running \textsc{MedianElimination} over the discretization.

\section{Main Technical Result}
In this section we present our algorithm and analyze its sample complexity, but before doing so let us first define the space of contracts $W$ that we can learn. The algorithm
is presented in \cref{sec:algorithm}.

\subsection{Learnable Contracts}
Let $w_0 > 0$ be a minimum wage for any outcome.
\begin{definition}[$B$-bounded contract]
A contract $w$ is $B$-bounded if $w_0 \le w(i) \le B$ for every $i \in [k]$.
\end{definition}

For a bounded contract, together with the assumption that the principal's profits are bounded, ensures that the principal's expected net profit $V(w)$ can be estimated statistically. 

\begin{definition}[Monotone-smooth contract]\label{def:monotone-smooth-contract}
A contract $w$ is monotone-smooth if for every $i \in [k-1]$ it holds that $0 \le w(i+1) - w(i) \le \pi(i+1) - \pi(i)$.
\end{definition}

For a monotone-smooth contract, \cref{eq:assump} ensures that, keeping the contract fixed, 
the principal's utility cannot decrease if the agent increases her effort level. In the 
sequel, this property allows us to bound the difference in the principal's utility between 
two similar contracts.

We define $W$ as follows.
\[
W = \left\lbrace w : w~\mbox{is monotone-smooth and $H$-bounded} \right\rbrace~.
\]
We are aware that this set seems restrictive at first glance, yet we argue that in some important special cases, the principal's optimal net profit is achieved by a contract from this set. For example, when there are only two outcomes or the utility of the agent is linear (see \cref{sec:apps}).

\subsection{Algorithm}
\label{sec:algorithm}
Let $w^\star \in W$ be an optimal contract in $W$, that is $V(w^\star) \ge V(w)$ for all $w \in W$. 
The goal of our algorithm is to find an \eps-optimal contract $w$, namely a contract for which $V(w^\star) \le V(w) + \epsilon$ within a predetermined number of rounds. We conjecture that it cannot be done in general therefore we make the following simplifying assumption.
\begin{assumption}[Bounded Risk-Aversion]
\label{assumption:rrn}
The agent's utility from wage $u$ is twice continuously-differentiable. Moreover, there exists a reference utility function $r$ that is a monotonically increasing, twice-differentiable, concave function such that $u''(w)/u'(w) \ge r''(w)/r'(w)$ for all $w > 0$. This is equivalent to $w \mapsto u'(w)/r'(w)$ being monotone-nondecreasing in $w$.
\end{assumption}
Intuitively, this assumption ensures that making small changes to a contract does not produce  behavior by the agent that is drastically different than if the agent's utility would have been $r$ instead of $u$.\footnote{An alternative, slightly less general, version of \cref{assumption:rrn} is: Assume there exists a finite $\eta>1$ such that $ -x u''(w)/u'(w)>\eta$ for all $w > 0.$ Note that the element on the left is the cannonical Arrow-Pratt relative risk aversion measure of the agent \cite{Arrow:65,Pratt1964}, and the element of the right corresponds with the Arrow-Pratt relative risk aversion measure of the Isoelastic utility function $r(w)=\frac{w^{\eta-1}-1}{\eta-1}.$}

Our algorithm works as follows. The principal initially constructs a cover $W_\eta$ of $W$, and then run \textsc{MedianElimination} on $W_\eta$. Indeed, the main technical difficulty in this paper is in defining $W_\eta$ properly so that the following result holds.
\begin{theorem}
\label{thm:discretization}
Suppose that \cref{assumption:fosd,assumption:breakties,assumption:rrn} hold. Let $\eta < r'(2H)\cdot H/k$. There exists a contract space $W_\eta$ such that for every contract $w \in W$, there is a contract $w' \in W_{\eta}$ for which $V(w) \le V(w') + k \eta / r'(2H)$. Moreover, the size of $W_{\eta}$ is at most $M = ((r(2H) - r(w_0)) / \eta)^k$, and $W_\eta$ can be constructed in time $O(M)$.
\end{theorem}

The proof of the theorem is found in \cref{sec:discproof}. Finally, we have our main result.
\begin{theorem}
\label{thm:main}
Suppose that \cref{assumption:fosd,assumption:breakties,assumption:rrn} hold. 
Let $\eta = \eps \cdot r'(2H)/2k$.
Executing \textsc{MedianElimination}$(\epsilon/2,\delta)$ on the set $W_{\eta}$ produces the following guarantee. With probability at least $1-\delta$, the algorithm outputs an $\eps$-optimal contract after 
\[
O \left( \left(\frac{4kH (r(2H)-r(w_0))}{\eps} \right)^{k+2} \cdot \log(1/\delta) \right)
\]
rounds.
\end{theorem}

\begin{proof}
By \cref{thm:discretization} and by the choice of $\eta$, there is a $w' \in W_{\eta}$ for which $V(w') \le V(w^\star) + \eps/2$.
By \cref{thm:me}, with probability $1-\delta$, \textsc{MedianElimination} returns a contract $\hat w \in W_{\eta}$ such that $V(\hat w) \le V(w') + \eps/2$. Combining both results we get
\[
V(\hat w) \le V(w')+\eps/2 \le V(w^\star) + \eps/2 + \eps/2 = V(w^\star) + \eps~,
\]
as required. Moreover, \textsc{MedianElimination} is done in the following number of rounds:
\[
    O \left( \frac{|W_{\eta}| H^2}{(\eps/2)^2} \log(1/\delta) \right) 
    = 
    O \left( \left(\frac{4kH (r(2H) - r(w_0))}{\eps} \right)^{k+2} \cdot \log(1/\delta) \right)
    ~.
    \qed
\]
\end{proof}

\subsection{Discretization of the Contract Space}
\label{sec:discproof}
In this section we prove \cref{thm:discretization}. We start by defining the notion of a coarse contract. To that end, we utilize the inverse function of $r$ (that exists everywhere since $r$ is increasing) which we denote by $r^{-1}$.

\begin{definition}[$\eta$-coarse contract]
A contract $W$ is $\eta$-coarse if there exists natural numbers $l_0,l_1,\ldots,l_{k-1}$ such that $w(1) = r^{-1}(r(w_0) + \eta \cdot l_0)$, and for $i \in [k-1]$, $w(i+1) = r^{-1}(r(w(i)) + \eta \cdot l_i)$.
\end{definition}
That is, in a coarse contract the ratios between wages of consecutive outcomes come from a discrete set of options. We define:
\[
W_\eta = \left\lbrace w : w~\mbox{is $\eta$-coarse and $2H$-bounded} \right\rbrace~.
\]
We prove the following.
\begin{lemma}
The size of $W_\eta$ is at most $M = ((r(2H) - r(w_0))/\eta)^k$. Moreover, $W_\eta$ can be constructed in time $O(k M)$.
\end{lemma}

\begin{proof}
The wage of outcome $i$ has the form $r^{-1}(r(w_0) + \eta \cdot l)$ for a natural number $l$, and satisfies $w(i) \le 2H$. Therefore, the number of choices for $w(i)$ is at most $(r(2H) - r(w_0))/\eta$. Since there are $k$ outcomes, there must be at most $M$ such contracts.
To construct $W_\eta$ we can go over all of its elements one-by-one, which takes $O(M)$ time.
\qed
\end{proof}

Finally let $w \in W$, we need to show that there is $w' \in W_\eta$ such that $V(w) \le V(w') + 2kH\eta$. We construct $w' \in W_\eta$ as follows.
We let $l_0 = \lceil (r(w(1) - r(w_0))/\eta \rceil$, and for $i \in [k-1]$, $l_i = \lceil (r(w(i+1)) - r(w(i)))/\eta \rceil$. Since $r^{-1}$ is also monotonically-increasing, we have that $w'(1) \ge w(1)$ as well as $r(w'(i+1)) - r(w'(i)) \ge r(w(i+1)) - r(w(i))$ for $i \in [k-1]$ by construction. Clearly, $w'$ is $\eta$-coarse and $w'(i) \ge w_0$, yet it remains to show that $w'(i) \le 2H$. For that, we have the following lemma.
\begin{lemma}
\label{lemma:discwageub}
We have for all $i \in [k-1]$, $r(w'(i)) \le r(w(i)) + \eta i$.
\end{lemma}
\begin{proof}
By construction, for each $i \in [k-1]$ we have $r(w'(i+1)) - r(w'(i)) \le r(w(i+1)) - r(w(i)) + \eta$. From this we entail that $r(w'(i)) - r(w'(1)) \le r(w(i)) - r(w(1)) + (i-1) \cdot \eta$. Since also by construction $r(w'(1)) \le r(w(1)) + \eta$, we get that $r(w'(i)) \le r(w(i)) + i \cdot \eta$ as required. \qed
\end{proof}
With the lemma at hand, and by assumption that $\eta < r'(2H)\cdot H/k$ we obtain
\[
    w'(k) 
    \le 
    r^{-1}(r(w(k)) + \eta \cdot k) 
    \le
    r^{-1}(r(H) + r'(2H) \cdot H)
    \le 
    2H~,
\]
using the concavity of $r$.
Therefore we have $w' \in W_\eta$.

We now show that, compared to $w$, under $w'$ the agent's effort 
cannot decrease. Then, we use this fact to bound the difference in the principal's utility between $w$ and $w'$.

In order to prove our first claim, we will make use of the following lemma.
\begin{lemma}[Grossman and Hart \cite{Grossman1983}]
\label{lem:G83}
Let $w^1$ and $w^2$ be contracts. Then,
\begin{align*}
\sum_{i=1}^k \left(f_{\hat e(w^1)}(i) - f_{\hat e(w^2)}(i)\right) \cdot \big(u(w^1(i)) - u(w^2(i)) \big) \ge 0.
\end{align*}
\end{lemma}

\begin{lemma}
\label{lem:lplus}
The effort level the agent chooses can only increase from $w$ to $w'$.
\end{lemma}
\begin{proof}
First, notice that since the wages only increase, had the agent accepted the contract $w$, 
i.e., chose an effort level different than 0, she would also accept contract $w'$. So, let 
$e' = \hat e(w')$ and $e = \hat e(w)$ be the effort levels the agent chooses under contracts $w'$ and $w$ respectively.

If we apply Lemma~\ref{lem:G83} with $w$ and $w'$, we obtain
\begin{align*}
\sum_{i=1}^k \big(f_{e'}(i) - f_{e}(i)\big)\cdot \left(u(w'(i)) - u(w(i)) \right) \ge 0~.
\end{align*}
Assume for now that $u(w'(i)) - u(w(i))$ is monotone nondecreasing in $i$. Given this, we will show by contradiction that $e' \succ e$. So, for the sake of contradiction assume that
$e' \prec e$. From the fact that $f_{e}$ dominates $f_{e'}$, \cref{eq:assump} implies that
\begin{align*}
\sum_{i=1}^k \big(f_{e'}(i) - f_{e}(i)\big) \cdot \big(u(w'(i)) - u(w(i)) \big) \le 0~.
\end{align*}
Thus,
\begin{align}
\label{eq:help}
\sum_{i=1}^k \left(f_{e'}(i) - f_{e}(i)\right) \cdot u(w(i)) = \sum_{i=1}^k \left(f_{e'}(i) - f_{e}(i)\right) \cdot u(w'(i)).
\end{align}
Therefore, by optimality of $e$ and $e'$ under contracts $w$ and $w'$ respectively, 
we obtain
\begin{align*}
& \sum_{i=1}^k \big(f_{e'}(i) - f_{e}(i)\big) \cdot u(w(i)) \le c(e') - c(e),~\text{and}~\\
& \sum_{i=1}^k \big(f_{e'}(i) - f_{e}(i)\big)\cdot u(w'(i)) \ge c(e') - c(e).
\end{align*}
Combining \cref{eq:help} with the two inequalities above, we obtain that 
\[
U(w',e') = \sum_{i=1}^k f_{e'}(i) \cdot u(w'(i)) - c(e') = \sum_{i=1}^k f_{e}(i) \cdot u(w'(i)) - c(e) = U(w',e)~.
\]
This means that the agent is indifferent between effort levels $e$ and $e'$ under contract 
$w'$. Since, by \cref{assumption:breakties}, the agent chooses the highest effort in this case, we must have 
$e \prec e'$ --- a contradiction.

Hence, in order to prove the lemma it suffices to prove that $u(w'(i)) - u(w(i))$ is monotone nondecreasing in $i$. 
This is equivalent to showing that $u(w'(i+1))-u(w'(i)) \ge u(w(i+1)) - u(w(i))$. 
Denote $c = r(w'(i)) - r(w(i))$. By construction, $r(w'(i+1)) - r(w(i+1)) \ge c$. 
Consequently, since $u$ is monotone nondecreasing, it suffices to show 
\[
    u(r^{-1}(r(w(i+1)) + c)) - u(r^{-1}(r(w(i)) + c)) 
    \ge 
    u(w(i+1)) - u(w(i)).
\]

Thus, the proof boils down to showing $c \mapsto u(r^{-1}(r(w(i+1)) + c)) - u(r^{-1}(r(w(i)) + c))$ is monotone nondecreasing in $c$. Taking the derivative with respect to $c$, we need to show
\[
    \frac{u'\bigl(r^{-1}(r(w(i+1)) + c)\bigr)}{r' \bigl(r^{-1}(r(w(i+1)) + c)\bigr)} 
    - 
    \frac{u'\bigl(r^{-1}(r(w(i)) + c)\bigr)}{r' 
    \bigl(r^{-1}(r(w(i)) + c) \bigr)}  
    \ge 
    0.
\]
However, the above holds since the agent is BRA (\cref{assumption:rrn}), and as $r^{-1}(r(w(i+1))+c) \ge r^{-1}(r(w(i))+c)$ due to both $r$ and $r^{-1}$ being monotone-increasing. \qed
\end{proof}

We can now bound the loss the principal suffers when she offers $w'$ instead of $w$.

\begin{lemma}
\label{lem:regret-bound}
It holds that $V(w) \leq V(w') + k\eta / r'(2H)$.
\end{lemma}
\begin{proof}
Since we focus on scenarios where the optimal contract is monotone-smooth, we get that the net profit of the principal at the optimal contract, $\pi(i) - w(i)$, is nondecreasing in $i$.
Furthermore, from \cref{eq:assump}, keeping $w$ fixed, the principal only benefits from an increase of the agent's effort level. Denote $e = \hat e(w)$ and $e' = \hat e(w')$. From \cref{lem:lplus} we know that  $e' \succ e$. We have,
\begin{align}
V(w) &= \sum_{i=1}^k f_{e}(i) \cdot \big(\pi(i) - w(i) \big) \nonumber \\
&\le \sum_{i=1}^k f_{e'}(i) \cdot \big( \pi(i) - w(i) \big) \nonumber \\
&= V(w') + \sum_{i=1}^k f_{e'}(i) \cdot  \big(w'(i) - w(i) \big).
\label{eq:principalutilitybound}
\end{align}
Now by \cref{lemma:discwageub}, $w'(i) \le r^{-1}(r(w(i)) + \eta \cdot i)$ for all $i \in [k]$. Note that $r^{-1}$ is convex. Then, as $\eta < r'(2H) \cdot H/k$, we obtain 
\begin{align*}
    w'(i) - w(i) 
    &\le 
    r^{-1}(r(w(i)) + \eta \cdot i) - w(i) \\
    &\le 
    \frac{\eta \cdot i}{r'(2H) \cdot H} \Bigl( r^{-1} \bigl(r(w(i)) + r'(2H)\cdot H \bigr) - w(i)\Bigr) \\
    &\le 
    \frac{\eta \cdot i}{r'(2H) \cdot H} \Bigl( r^{-1} \bigl(r(H) + r'(2H)\cdot H \bigr) - H \Bigr) \\
    &\le 
    \frac{\eta \cdot i}{r'(2H) \cdot H} \bigl( 2H - H \bigr) \\
    &\le
    \frac{\eta \cdot k}{r'(2H)},
\end{align*}
where the second and third inequalities are by the convexity of $r^{-1}$, and the fourth inequality is by the concavity of $r$.
Combining the latter with \cref{eq:principalutilitybound}, we get 
$V(w) \le V(w') + \eta \cdot k / r'(2H)$.
\qed
\end{proof}

%


\section{Applications}
\label{sec:apps}
In this section we highlight two cases that received attention in the past. For each of 
them, when \cref{assumption:fosd,assumption:breakties,assumption:rrn} hold, the optimal 
contract will be in  the set $W$, and thus by learning an $\varepsilon-$optimal contract in 
$W$, we approximate the best contract the principle could have offered the agent had she known her utility function, effort levels and costs, and the 
distributions they induce over outcomes.

\paragraph{\bf Two outcomes.}
Firstly, we focus on the case where there are only two outcomes and show that the optimal contract is in $W.$
\begin{lemma}
When there are only two outcomes, the optimal contract is monotone-smooth and $2H$-bounded.
\end{lemma}
\begin{proof}
By \cite{Grossman1983}, the optimal contract in the two outcome case is of the shape: $w(2)=w(1)+a(\pi(2)-\pi(1))$ for some $a\in[0,1]$. By plugging this expression into the inequality in \cref{def:monotone-smooth-contract}, we get that the optimal contract is monotone-smooth. To see that the optimal contract is $2H$-bounded, let $e$ denote her chosen effort level. Since the principal's utility at the optimal contract is nonnegative, $f_e(1) w(1) + f_e(2) w(2) \le f_e(1) \pi(1) + f_e(2) \pi(2) \le H$, and in particular $w(1) \le H$. Now, $w(2) = w(1) + a(\pi(2)-\pi(1)) \le H + 1\cdot H = 2H$. \qed
\end{proof}

Thus, by \cref{thm:main}, applied to our discretized contract space, \textsc{MedianElimination} finds an \eps-optimal contract under \cref{assumption:fosd,assumption:breakties,assumption:rrn}.
Note, the FOSD assumption (\cref{assumption:fosd}) is standard in the literature, hence our result essentially 
requires only the bounded risk-averse assumption (\cref{assumption:rrn}). 


\paragraph{\bf Risk neutral agent.}
\cite{Carroll2015} studies a setting in the agent is risk-neutral, and the principal has only partial knowledge of the agent's action space. Had the principal known 
the complete action space of the agent, the optimal contract would have been linear. 
In this setting, the principal can derive the optimal \emph{linear} contract  with respect to only the known actions of the agent. \cite{Carroll2015} show that her profit from the actual action taken by the agent (which can be one that the principal is unaware of) can only be higher than the principal's expectations.

In the following lemma we show that when the agent is risk-neutral, the optimal theoretical contract is in $W$, and thus our algorithm learns a contract that approximates it.
\begin{lemma}
If the agent is risk-neutral then the optimal contract is $H$-bounded and monotone-smooth.  
\end{lemma}
\begin{proof}
When the agent is risk neutral, the optimal contract is of the shape $w(i)=\pi(i) - \alpha$ for some constant $\alpha \in \mathbb{R}_{+}$ (Proposition 14.B.2 on page 482 in \cite{Mas-Colell1995}). As $\pi(i+1)>\pi(i),$ this contract is monotone-smooth. And as $\pi(k) \le H$ it is also $H$-bounded. \qed
\end{proof}

\section{Conclusions}
In this paper we studied the principal-agent problem when the principal has zero information about the the agent. We introduced the class of monotone-smooth contracts and showed that when the optimal contract is monotone-smooth and the agent is bounded
risk-averse, then we can learn an approximately optimal contract. We complemented this result with a multi-armed bandit algorithm that finds an approximately optimal contract and we provided bounds on the number of samples it needs. Then, we applied
our algorithm to two fundamental scenarios. The first one is when the output space 
of the task is binary and the second one is when the agent is risk-neutral. 
Economic theory suggests that the optimal contract \emph{is} monotone-smooth. 
Thus the net  profit of the principal generated by the resulting contract 
approximates the optimal net profit she can achieve in general. To the best of our
knowledge these are the first positive results even regarding approximately optimal
contracts. 

Several intriguing questions remain. It is interesting to understand whether the assumption of bounded risk-aversion is needed to guarantee learning of monotone-smooth contracts. The answer to this question is not obvious even when there are only two outcomes. Furthermore, we wish to find other conditions and assumptions that allow learning. In our model, we assume that the agents are identical. Can we learn $\eps$-optimal contracts if there are many different types of agents? We conjecture that we {\em can} under the suitable assumptions, like the setting of~\cite{Ho2016}. On the other hand, we furthermore conjecture that a there exist cases in which learning is {\em not} possible at all. Lower bounds, or even partial characterizations of such cases would be of great interest.


\bibliographystyle{splncs04}
\bibliography{contracts}

\appendix

\end{document}